\documentclass[article,reqno]{amsart}
\usepackage{graphicx}
\usepackage{color}
\usepackage{amsmath,amssymb}

\newcommand{\eqn}{\begin{eqnarray}}
\newcommand{\feqn}{\end{eqnarray}}
\newcommand{\beq}{\begin{equation}}
\newcommand{\eeq}{\end{equation}}
\newcommand{\bes}{\begin{equation*}}
\newcommand{\ees}{\end{equation*}}
\newcommand{\smat}{\left( \begin{smallmatrix}}
\newcommand{\smct}{\end{smallmatrix}\right)}

\newcommand{\beqnl}{\begin{eqnarray}}
\newcommand{\eeqnl}{\end{eqnarray}}

\begin{document}

\title{Tunneling method for Hawking radiation in the Nariai case}
\author{F.~Belgiorno, S.L.~Cacciatori, F.~Dalla Piazza}

\address{Francesco Belgiorno, 
Dipartimento di Matematica, Politecnico di Milano, Piazza Leonardo 32, 20133 Milano, Italy, and INdAM-GNFM, Roma, Italy, and INFN, Milano, Italy}
\email{francesco.belgiorno@polimi.it}
\address{
Sergio L. Cacciatori, 
Department of Science and High Technology, Universit\`a dell'Insubria, Via Valleggio 11, IT-22100 Como, Italy
INFN sezione di Milano, via Celoria 16, IT-20133 Milano, Italy}
\email{sergio.cacciatori@uninsubria.it}
\address{
Francesco Dalla Piazza, 
Department of Science and High Technology, Universit\`a dell'Insubria, Via Valleggio 11, IT-22100 Como, Italy
INFN sezione di Milano, via Celoria 16, IT-20133 Milano, Italy}
\email{fdallapiazza@gmail.com}

\date{Received: date / Accepted: date}
\maketitle

\begin{abstract}
We revisit the tunneling picture for the Hawking effect in light of the charged Nariai manifold, 
because this general relativistic solution,    
which displays two horizons, provides the bonus to allow the 
knowledge of exact solutions  of the field equations. We first perform a 
revisitation of the tunneling ansatz in the framework of particle creation in external fields 
{\sl \`a la} Nikishov, which corroborates the interpretation of the semiclassical emission rate $\Gamma_{emission}$ 
 as the conditional probability rate for the creation of 
a couple of particles from the vacuum. 
Then, particle creation associated with the Hawking effect on the Nariai manifold is calculated in two ways.  
On the one hand, we apply the Hamilton-Jacobi formalism for tunneling, in the case 
of a charged scalar field on the given background. 
On the other hand,  the knowledge of the exact solutions for the 
Klein-Gordon equations on Nariai manifold, and their analytic properties on the extended 
manifold, allow us a direct computation of the flux 
of particles leaving the horizon, and,  
as a consequence, we obtain a further corroboration of the semiclassical tunneling picture from the side of S-matrix formalism. 
\end{abstract}

\maketitle


\section{Introduction}

Tunneling through the horizon is a longstanding approach to Hawking effect since the 
seminal papers by S.W.Hawking \cite{haw-cmp,hartle-hawking}. 
Between the various methods concurring in corroborating 
the original calculations, the so-called tunneling method has been proposed.  We limit 
ourselves herein to quote some seminal papers and a fine review 
\cite{damour-ruffini,srinivasan,visser-essentials,parikh-prl,singleton-tunnel,vanzo-rev}.  
The Hamilton-Jacobi (HJ henceforth) formalism for the calculation of the 
particle creation associated with the Hawking effect represents a 
semiclassical approach where  the classical action of particles is 
computed along trajectories which pass through the horizon. 
A special version of the method is represented by 
the Parikh-Wilczek approach \cite{parikh-prl}, where a  
tunneling through the horizon of a particle arises   
because of a quite unexpected mechanism, where the tunneling particle sets up the barrier by 
energy conservation, as nicely described by Parikh \cite{parikh-tunnel}.\\ 

We revisit the HJ tunneling method for the Hawking effect by taking into account a 
the charged Nariai solution, in view of the fact that it allows to gain the knowledge of exact solutions  
of the field equations even in the inner black hole region. 
This is our basic reason for studying tunneling in this particular manifold, 
with the aim of considering it as a further benchmark for the tunneling method, which is of course a very 
useful and simple method for deriving the Hawking effect, but whose status is not so firmly grounded 
on the theoretical side \cite{moretti-pinamonti}. With this aim, we first reinterpret the 
semiclassical emission rate
\begin{equation}
\Gamma_{emission} =\exp \left(-2 \mathrm{Im} S\right),
\label{eq-gamma}
\end{equation}
{\color{black} where $S$ is the classical action}, 
as the conditional probability rate for the creation of 
a couple of particles from the vacuum (to be intended as the state of absence of particles). 
{\color{black} We remark that eq. (\ref{eq-gamma}) is the standard reference equation for the 
literature on the tunneling method, and e.g. in static backgrounds one may separate in the full action $S$ a 
temporal part $\omega t$, where $\omega$ is the particle energy,  and a spatial one $S_0 (x)$ which depends only on the spatial variables, and 
eq. (\ref{eq-gamma}) is often written as $\Gamma_{emission} =\exp \left(-2 \mathrm{Im} S_0\right)=\exp \left(-2 \mathrm{Im} \int p dq\right)$, where $p,q$ are canonically conjugate (see e.g. \cite{singleton-tunnel}). 
The latter expression, as well as (\ref{cond-gamma}), has the drawback to be not invariant under canonical transformations, as 
is remarked first in \cite{chowdhury} and then in \cite{singleton-tunnel,singleton-background}. Such an invariance is achieved by the expression 
$
\Gamma_{emission} =\exp \left(-\mathrm{Im} \oint p dq \right).
$
See also sect. \ref{hj-nariai} for further discussion.}\\
Then, we consider the field equations on the Nariai manifold, and set up a scattering picture 
for the tunneling process, whose key-ingredient is the requirement for analyticity of the 
exact solutions both in the inner region and in the outer one with respect to the black 
hole horizon. Explicit computation of the flux of particles through the horizon corroborates 
the standard tunneling ansatz, which is also taken into account. 

The present analysis completes our previous 
studies concerning the quantum instability of the charged Nariai solutions \cite{belcacciadalla-fermions,belcacciadalla-bosons}, where the Hawking effect was not derived. Moreover, we mention that in \cite{KSS} an early study about the instability due to quantum matter of Nariai-like metrics appeared.

\section{The HJ method for tunneling} 

The basic idea of the HJ method is very simple, and consists in adopting the WKB approximation 
and computing the tunneling probability for a straddling mode, to be intended as a mode 
whose wave function is regular on the horizon and also defined across the horizon itself 
\cite{damour-ruffini,visser-essentials,massar-parentani}. Subtleties occur if singular 
coordinate systems on the horizon are adopted, as pointed out e.g. in \cite{singleton-tunnel,singleton-background,singleton-subtleties,singleton-wkb,singleton-unruh,vanzo,vanzo-rev}. 
Former calculations appear in \cite{srinivasan}, and further development are contained in 
\cite{shankaranarayanan-cqg,shankaranarayanan-mpla}. 
A thoroughful analysis and review is contained in \cite{vanzo-rev}, to which we refer the reader for a 
more complete list of references. We also refer to \cite{srinivasan,shankaranarayanan-cqg,shankaranarayanan-mpla,kerner}. 

As a basic ingredient of the approach, 
we have the classical action $S$ of particles (massless or not), to be computed along trajectories 
which pass through the horizon.  The semiclassical emission rate is given by 
(\ref{eq-gamma}), whose 
right-hand side is easily realized to correspond to the standard form for the rate of emission 
associated with tunneling through a potential barrier in the WKB approximation. Still, 
the barrier is non-standard, being present a single turning point against the usual couple 
of turning points for standard barriers.  The horizon plays the role of a unattainable limiting region for signals in the inner 
region of the black hole, much more than a real potential barrier. 
Moreover, a very non-trivial transition between a spacetime region 
with a time-dependent metric (black-hole region) to a static region (exterior of the black hole) 
is being occurring, so it is the case to remark that `standard interpretations' are not so 
well grounded or, at the very least, free of misinterpretations. See also \cite{moretti-pinamonti}. 
By following \cite{vanzo}, it is interesting to write down the action as follows:
\begin{equation}
S =\int dS =\int_\gamma (\partial_{x^i} S) dx^i,
\end{equation}
where $dS$ is the one-form corresponding to the differential of $S$, and an integration 
along {\color{black} an oriented, null path is understood, and this is at the root of the so called null geodesic method 
\cite{vanzo-rev}}. In such a way, $dS$ is written in terms of the differential 
of coordinates times the conjugate momenta $p_i=(\partial_{x^i} S)$ for $i=0,1,2,3$ (a change of sign 
in the 0-component can occur with respect to this definition). 

\section{Tunneling method and a trick \`a la Nikishov}

Our ansatz herein is that the probability rate of pair creation near the black hole 
horizon $\Gamma_{emission}$ can be interpreted as the conditional probability rate for the creation of 
a couple of particles from the vacuum (to be intended as the state of absence of particles). This 
interpretation is non-standard, and suggests that $\Gamma_{emission}$ 
is just more the square of the relative weight between the outer part and the inner 
part of the straddling mode than the pair-creation rate itself. This argument is to be compared with the 
argument in \cite{sannan}, which is relative to the original picture by Damour and Ruffini \cite{damour-ruffini}. \\
For the 
following general picture, we refer to \cite{nikishov,damour-mg1}. We recall that 
the imaginary part of the effective action $W$ is the signal of particle production. Indeed, 
the permanence of the vacuum has probability $<1$: particle creation occurs with probability (per 
unit time) 
\begin{equation}
P_{0_{in}\to 0_{out}} =\exp \left(-2 \mathrm{Im} W\right).
\label{vacuum-persistence}
\end{equation}
One can notice the resemblance with the formula defining $\Gamma_{emission}$, but the relation 
between $\mathrm{Im} W$ and $\Gamma_{emission}$ is not so straightforward. Still, it exists and is 
found below.\\
Basically, the following idea is pursued. We proceed as in \cite{damour-mg1} 
for the general picture.\\  
Let us introduce, for a diagonal scattering process,
\begin{equation}
n_i^{IN}= R_i n^{OUT}_i + T_i p^{OUT}_i,
\end{equation}
where $n_i$ stays for a negative energy mode and $p_i$ for a positive energy one. In case an 
inner product different from the standard one for bosonic and fermionic fields occurs, `positive 
energy' should be replaced by `positive norm' (and analogously for negative energy). 
$p_i^{IN},n_i^{IN}$ form a scattering basis for the IN states, and $p^{OUT}_i,n^{OUT}_i$ 
form a scattering basis for the OUT states. 
$T_i$ is the transmission coefficient and $R_i$ is the reflection one. It is evident that above we 
have written a Bogoliubov transformation between IN and OUT states, so the following identification 
is also true: 
$
R_i = \alpha_i,\ 
T_i = \beta_i.
$
Moreover, one defines as in \cite{damour-mg1}
\begin{equation}
\eta_i:=|T_i|^2,
\end{equation}
which can be shown to coincide with the mean number per unit time and unit volume
of created particles. One has 
$
|R_i|^2 = 1\mp \eta_i,
$
where, here and in the sequel, the upper sign holds for fermions and the lower one for bosons. 
By interpreting {\sl \`a la} Stueckelberg the scattering process, one can also obtain
$
n^{OUT}_i= R_i^{-1} n_i^{IN} - R_i^{-1} T_i p^{OUT}_i,
$
which is interpreted as the scattering of a negative mode incident from the future
and which
is in part refracted in the past and in part reflected in the future. The new
reflection amplitude $- R_i^{-1} T_i$ is such that the reflection coefficient
\begin{equation}
|R_i^{-1} T_i|^2 = \frac{\eta_i}{1\mp \eta_i}=\tilde{P} _i (1|0)
\end{equation}
can be interpreted as the conditional probability rate $\tilde{P} _i (1|0)$ for the creation of the pair
$n^{OUT}_i,p^{OUT}_i$, starting from absence of particles in that state. 
The conditional probability rate for $n$ couples is 
$
\tilde{P} _i (n|0)=(\tilde{P} _i (1|0))^n.
$
Of course, in the fermionic case only $n=1$ is allowed. 
The probability rate for $n$ couples is
$
\tilde{P}_i (n)=\tilde{P}_i (n|0) \tilde{P}_i (0).
$
$\tilde{P}_i (0)$ represents the probability rate that no particles are 
created in the given state $i$. It can be calculated as 
follows: 
$
\sum_n \tilde{P}_i (n)=1 =\tilde{P}_i (0) \sum_n \tilde{P}_i (n|0) ,
$
and then 
\begin{equation}
\tilde{P}_i (0)= \left(1\mp \eta_i\right)^{\pm 1}. 
\end{equation}
The persistence of the vacuum is given by
$
P_{0_{in}\to 0_{out}} =\prod_i \tilde{P}_i (0).
$
Then we can infer 
\begin{equation}
2 \mathrm{Im} W=\mp \sum_i  \log (1\mp \eta_i).
\label{im-effact}
\end{equation}
As to the mean number of created couples, we have 
\begin{equation}
<n_i>=\sum_n n \tilde{P}_i (n)=\eta_i = |\beta_i|^2.
\end{equation}
As a consequence, in the above formulas we realize that $\eta_i \mapsto <n_i>$ is allowed.

Let us apply the above picture to our specific case. We interpret $\Gamma_{emission}$ as follows: 
\begin{equation}
\Gamma_{emission}= \tilde{P} _{\omega}(1|0),
\label{cond-gamma}
\end{equation}
where $\omega$ identifies the quantum state. In the present case, we get for bosons 
$
\tilde{P}_{\omega}(0)=1- \exp \left(- \beta \omega\right),
$
with $\beta=\beta_H=1/(k_{boltzmann} T_H$) ($T_H$ is the black hole temperature). As a consequence, one gets 
$
\tilde{P}_{\omega}(n)=(1- \exp \left(- \beta \omega\right)) \exp \left(- \beta \omega n\right).
$
It is then easy to show that the mean number of created pairs in the state with energy $\omega$ is 
\begin{equation}
<n_\omega>=\sum_{n=0}^\infty n \exp \left(- \beta \omega n\right)= \frac{1}{\exp \left(\beta \omega\right) -1},
\end{equation}
which is the correct result. This argument is substantially equivalent to the one of ref. \cite{kim}. Note that a thermal 
particle distribution is obtained without recurring to detailed balance arguments.\\
It is also worth noticing that it holds
\begin{equation}
\mathrm{Im} W = -\frac{1}{2} \int d\omega\; \log (1-\exp (-\beta \omega)),
\label{im-w}
\end{equation}
which is the expected result\footnote{{\color{black} For the sake of completeness, one should write 
$\mathrm{Im} W = \frac{1}{2} \sum_{\omega,l,m} \log (1+<n_{\omega,l,m}>)$, which takes into account 
the full dependence on quantum numbers, and one realizes that the label $\omega$ introduced in (\ref{cond-gamma}) is split, with some abuse of language, 
into $\omega,l,m$, where $\omega$ is the energy, and $l,m$ are the usual 
quantum numbers for angular momentum.}}  \cite{kim,stephens}. The calculation in the fermionic case is analogous, and is based 
on the fact that the WKB approximation for the Dirac equation coincides with the HJ equation. The only change 
is the statistics. 
{\color{black} We point out again the substantial difference between the expressions for $\Gamma_{emission}$ and 
$\exp \left(-2 \mathrm{Im} W\right)$ appearing in (\ref{vacuum-persistence}). As shown above, $\Gamma_{emission}$ 
is the conditional probability (\ref{cond-gamma}) for the emission of a pair labeled by  $\omega$, whereas  
$\exp \left(-2 \mathrm{Im} W\right)$ is the probability that, for any field mode with label $\omega$, 
 there is not a quantum instability in the field at hand, 
and then it involves a sum over all values of $\omega$ (cf. (\ref{im-effact})).}\\
Extensions to the cases where one takes into account also the backscattering (which is 
mandatory in 4D) are discussed in \cite{sannan}. We also notice that the above interpretation concerning the 
meaning of $\Gamma_{emission}$ hold true also for the Parikh-Wilczek approach.

\section{Charged Nariai manifold}
\label{sec-nariai}

We describe herein the electrically charged Nariai solution. 
We shall consider Kruskal-like coordinates, that are introduced in the following 
for the black hole horizon $\chi=\pi$, which is our main focus, and then for the cosmological 
horizon $\chi=0$ (see below).\\
The manifold is described by the metric \cite{romans,mann,bousso}
\beq
ds^2 = \frac{1}{A} (-\sin^2 (\chi) d \psi^2 + d\chi^2) + \frac{1}{B} (d\theta^2+
\sin^2 (\theta) d\phi^2),
\label{nariai-metric}
\eeq
with $\psi \in {\mathbb{R}}, \chi\in (0,\pi)$, and the constants
$B=\frac{1}{2 Q^2}\left( 1-\sqrt{1-12 \frac{Q^2}{L^2}} \right)$, $A=\frac{6}{L^2}-B$
are such that $\frac{A}{B}<1$, and $L^2:= \frac{3}{\Lambda}$. The black hole horizon
occurs at $\chi=\pi$. This manifold has finite spatial section.
In the Euclidean version,
it corresponds to two spheres characterized by different radii.
For the gauge potential we can choose
$A_{i} = -Q \frac{B}{A} \cos (\chi) \delta_{i}^0$.\\
We consider the $\psi-\chi$ part of the metric, and introduce a diffeomorphism   
$\chi=\chi(r)$ such that
$
\frac{1}{A} (-\sin^2 (\chi) d \psi^2 + d\chi^2) = f(r) \frac{1}{A} (- d \psi^2 + dr^2).
$
This can be obtained for 
$
f(r)=\sin^2 (\chi)=\frac{1}{\cosh^2 (r)},
$
and we can choose the branch 
$
\chi=2 \arctan (\exp (-r)),
$
which is useful because the tortoise-like coordinate $r$ with this choice is 
such that $r\to -\infty$ as $\chi \to \pi^-$, as e.g. in the Schwarzschild case. Note that 
we get $\cos (\chi)=\tanh (r)$.\\ 
We need to introduce Kruskal-like coordinates. 
Then, we first 
introduce null coordinates
\eqn
u &=& \frac{1}{\kappa} (\psi-r),\\
v &=& \frac{1}{\kappa} (\psi+r),
\feqn
where $\kappa=\sqrt{A}$ is the surface gravity;  
then we can define the Kruskal-like coordinates adapted to the black hole 
horizon region $\chi=\pi$:
\eqn
U &=&- \exp (-\kappa u),\\
V &=& \exp (\kappa v).
\feqn
Then we obtain
$
ds^2=-\frac{4}{A} \frac{1}{1-UV} dU dV + \frac{1}{B} (d\theta^2+
\sin^2 (\theta) d\phi^2).
$
We note that in our latter coordinate chart we get
$
\partial_\psi=-U \partial_U + V \partial_V,
$
which will be useful in the following. 
We also need to introduce a gauge transformation in order to obtain a 
gauge potential which is regular on the horizon:
$
A'_\mu=A_\mu - \partial_\mu G, 
$
where
$
G=\frac{eQB}{A} \psi.
$
Then we get
$
e A'_U=\frac{-1}{2U} e A'_0 =-\frac{eQB}{ A} \frac{V}{1-UV}.
$
Analogously, we have
$
e A'_V=\frac{1}{2 V} e A'_0 =\frac{eQB}{ A} \frac{U}{1-UV}.
$
The aforementioned gauge transformation is such that the following shift 
in the one-particle energy occurs:
\beq
\omega\mapsto \omega+e \Phi_H,
\eeq
where
$
e \Phi_H=\frac{eQB}{A}.
$
For the cosmological horizon region we introduce further Kruskal 
coordinates 
$
\bar{U} = \exp (\kappa u),\ 
\bar{V} = -\exp (-\kappa v).
$ 
The cosmological horizon $\chi=0$ corresponds to $\bar{U}=0$ and $\bar{V}=0$, 
and analogous equations can be found. In particular, a further gauge transformation 
regularizing the potential on the cosmological horizon can be analogously given.

\section{Hawking effect on Nariai manifold in the HJ formalism}
\label{hj-nariai}

In this section, we set up the HJ approach to tunneling for the Nariai charged solution in the case of a charged scalar field. 
It is worth mentioning that Hawking effect in the tunneling framework in de Sitter spacetime has been considered several 
times, in different situations which are mainly involved with the Schwarzschild-de Sitter solutions. See e.g. refs. 
\cite{medved,parikh-ds,zhang,kim-ds,farmany,rahman}, where both the HJ approach and the Parikh-Wilczek one are considered.\\ 
We focus our attention on the HJ equation for the black hole Kruskal patch (we recall that the black hole horizon corresponds to 
$\chi=\pi$ in the original coordinates (\ref{nariai-metric})): 
\beq
- (1-UV)^2 (k_U+e A_U)(k_V+e A_V)+\mu_l^2=0,
\eeq
where we define $\mu_l^2:=\frac {\mu^2}A +\frac BA l(l+1)$; furthermore, we have  
$k_U=\partial_U S $, $k_V=\partial_V S$, 
and $S$ is the action. We also take into account that, 
due to separation of variables, 
\beq
S = -\frac{\omega}{\kappa} \psi+h(r)+B l(l+1),
\eeq
and then {\color{black} the action is automatically separated in its temporal part and in its spatial one. 
As a consequence,} we have
$
\partial_\psi S =- U k_U+V k_V =-\frac{\omega}{\kappa}.
$
The fundamental amplitude to be calculated is 
\beq
\Gamma = \exp( -2\; \mathrm{Im}  \int dU (\partial_U S+e A'_U)).
\label{cond-una}
\eeq
{\color{black} We notice that the former expression (\ref{cond-una}) implicitly takes into account the 
temporal contribution to $\exp \left(-\mathrm{Im} \oint p dq \right)$ one has to include in order to obtain a consistent 
implementation of the tunneling picture \cite{singleton-subtleties,singleton-wkb,singleton-unruh}. Explicitly, 
one has $\Gamma=\exp( -\mathrm{Im} \left( \omega \Delta t_{out}+\omega \Delta t_{in}-(\int p^{out} dq-\int p^{in} dq) \right) )$, where 
$\Delta t_{out}$ refers to the temporal contribution for outgoing particles, and analogously for $\Delta t_{in}$ \cite{singleton-subtleties}.  
See also \cite{degill-unruh} for further discussion. The point is that 
we are substantially implementing the null geodesic method discussed in \cite{vanzo,vanzo-rev}. 
The result we obtain is thus correct, because, as 
shown in \cite{vanzo-rev} (see in particular sect. 3 therein), the null geodesic method is 
covariant and invariant under canonical transformations, and equivalent to the HJ method.} 
In order to perform the above integration, we must regularize the integral by choosing 
a suitable circuit in the complex plane. This 
amounts to a definition of the aforementioned integral, which is otherwise ill-defined.
We obtain
{\color{black}
\eqn
&&\int_\gamma\; dU\; K_U(U)= \int_{U_1}^{\epsilon}\; dU\;  K_U (U)\cr
&&+\int_{C_{\epsilon}}\; dz\;  K_U (z)
+\int_{-\epsilon}^{U_2}\; dU\;  K_U (U),
\feqn
where $U_1>0,U_2<0$, $K_U:=\partial_U S+e A'_U $, and $C_{\epsilon}$ is a semicircle which is oriented anti-clockwise centered in $U=0$ and in the upper half $U$-plane, and then an antiparticle state is occurring, in agreeement with the picture in \cite{damour-ruffini,visser-essentials}. See also the following section}. The only contribution to 
the imaginary part of the action $S$ is due to $\int_{C_{-\epsilon}}\; dz\;  K_U (z)$. Indeed,  
a simple application of the fractional residue theorem (\cite{gamelin}, p. 209) leads to 
\begin{equation}
\lim_{\epsilon \to 0^+}\int_{C_{\pm \epsilon}}\; dz\;  K_U (z) =\pm  i \pi \mathrm{Res} [K_U(z),U=0],
\end{equation}
{\color{black} where $C_{\pm \epsilon}$ refers to a semicircle oriented clockwise for negative sign ($C_{-\epsilon}$) and anti-clockwise for 
positive sign $(C_{+\epsilon}$),}  
which in the present case gives 
\beq
2\; \mathrm{Im}  \int dU (\partial_U S+e A'_U)=\frac{2\pi}{\kappa} \left(\omega-\frac{eQB}{A}\right),
\eeq
which is the expected result. Absorption occurs instead with conditional probability equal to 1, which is both the classical result and also 
compatible with the detailed balance argument. 
In the case of the cosmological horizon, analogously one finds
\beq
2\; \mathrm{Im}  \int d\bar{U} (\partial_{\bar{U}} S+e A'_{\bar{U}})=\frac{2\pi}{\kappa} 
\left(\omega+\frac{eQB}{A}\right).
\eeq
\hphantom{ccc}

\section{Hawking emission in S-matrix formalism}

The semiclassical picture represented by the tunneling ansatz can also be corroborated by an S-matrix approach, due to the fact that 
we can calculate exact solutions of the Klein-Gordon equation and know their analyticity properties 
on the Nariai manifold. 
We consider the Hawking effect from the point of view of scattering theory. In order to set up a tunneling 
picture, we need to consider a rather unusual picture where the scattering takes place part inside the 
black hole horizon and part outside in the external region, with the black hole horizon playing the role 
of barrier. Needless to say, this is the original picture proposed by Hawking and then by Hartle and Hawking 
in a path integral formalism \cite{haw-cmp,hartle-hawking}. Herein, we simply limit ourselves to propose this 
picture for the case at hand, in a non-dynamical situation.\\
We start from the Klein-Gordon equation of the Nariai background  for the scalar field $\Phi$ \cite{belcacciadalla-bosons}. 
We use rescaled physical quantities, for example we have for particle energy 
$
\omega=\frac{\omega_{phys}}{\kappa}.
$
In the following, we shall indicate with $\omega$ both the rescaled variable and the physical value, in order to avoid 
to make heavy the notation.  
    Separation of variables $\Phi = e^{-i \omega \psi} Y_{lm}(\Omega) \Psi(\chi)$ and a change of variable $t=-\cos \chi$ lead to the following reduced `radial' equation  \cite{belcacciadalla-bosons}
\beq
(1-t^2)\Psi''-2t\Psi'+\left[\frac 1{1-t^2} (\omega +e Q\frac BA t )^2 -\mu_l^2  \right]\Psi=0,
\eeq
where the prime is the derivation w.r.t. $t$. Note that this equation is invariant under $\{t\rightarrow -t,\ Q\rightarrow -Q\}$ so that
we can look at the singularity in $t=1$ only and obtain the properties of the singularity in $t=-1$ by $Q\rightarrow -Q$. Now, the behaviour 
of the above equation near $t=1$ suggests to set
\begin{eqnarray}
 \Psi(t)=(1-t)^{l_+} (1+t)^{l_-} \Phi(t),
 \end{eqnarray}
where $l_\pm= \frac i2 |\omega \pm e E |$, and 
$E:=Q\frac{B}{A}$. 
The equation for the function $\Phi$ is
$$
(1-t^2)\Phi''-2(t-l_+ (1-t)+l_- (1+t))\Phi' -d_l\Phi=0,
$$
where $d_l:=\mu_l^2-\omega^2+l_+ +l_- -(l_+-l_-)^2$. 
It is easy to infer that the general solution for $\Phi$ is given by a hypergeometric function:  
\beq
\Phi(t)=C_+ F (a,b,c_+,t_+)+C_- F (a,b,c_-,t_-),
\label{ipergeometrica}
\eeq
where $F(a,b;c;z)$ is the usual hypergeometric function and $a:=i e E+\frac 12+i\sqrt{\Delta}, b:=i e E+\frac 12-i\sqrt{\Delta},
 c_\pm:=i(e E\pm\omega)+1, t_\pm:=\frac {1\mp t}2$, and 
$
\Delta:=\mu_l^2+(eE)^2-\frac{1}{4}.
$
The above solution holds in the level-crossing region, 
i.e. for $-eE<\omega<eE$, where also pair emission of charged particles occurs \cite{belcacciadalla-bosons}. 
Still, that solution is easily shown to hold also outside the level crossing region. 
We then consider Kruskal-like coordinates as in the previous section and then 
we get $\psi(U,V)$ and $t(U,V)$ (which are not explicitly calculated). Of course, 
using $U,V$ one is allowed to extend solutions inside the black hole. Analytic continuation 
is allowed, and we have to look about the branch singularities of the hypergeometric functions combined with the one
associated to the factor
$
(1-t)^{l_+} (1+t)^{l_-}.
$
We can easily deduce that $\Psi(t)$ presents the same singularities as the function\footnote{To be precise this is true only at finite, since they have different singularities at infinity.}
$
\tilde \Psi(t):=C_1(1-t)^{l_+} (1+t)^{-l_-}+C_2(1-t)^{-l_+} (1+t)^{l_-}.
$
Thus, we have a logarithmic branch point at $t=1$ (black hole 
horizon) and one at $t=-1$ (cosmological horizon). This implies the presence of a logarithmic 
branch cut (we choose the negative real axis, as usual) and the appearance of a suitable 
exponential factor as one passes the horizon. See below.\\
In a scattering picture, one sets up a so-called straddling mode \cite{visser-essentials}. 
This mode can be obtained by analytic continuation from the outgoing one as in the original 
picture by Damour and Ruffini \cite{damour-ruffini}. See also \cite{bezerra,vieira} for more 
recent applications of the Damour-Ruffini picture.\\
The most simple analysis can be performed in the case of a uncharged massless scalar field $e=0,m=0$ 
in the s-wave $l=0$ (which is the leading contribution to the Hawking radiation). One obtains 
for the $\psi-t$ part $\eta (\psi,t)$ of the wave function in Kruskal-like coordinates 
\beq
\eta(U,V)=c_1 (-U)^{i \omega} 
+ c_2 V^{-i \omega} 
\label{sol-gen}
\eeq
which gives us an outgoing mode emerging from the black hole horizon for $c_2=0$. Then we 
find 
$
\eta(U,V)_{outgoing}=c_1 (-U)^{i \omega}; 
$
a negative norm mode which straddles the horizon is obtained as follows \cite{damour-ruffini,visser-essentials}:
\beq
\eta(U,V)_{straddle}=N_s \eta(-U+i\epsilon,V)_{outgoing},
\label{straddle}
\eeq
by analytic continuation. In passing the black hole horizon $U=0$ ($r=-\infty$) a contribution 
from the logarithmic branch point arises, so that 
\beq
\eta(U,V)_{straddle}=N_s \left\{ 
\begin{array}{ll}
(-U)^{i \omega} e^{-\pi \omega} & -U>0,\cr
U^{i \omega} & U>0.
\end{array} \right.
\eeq
As a consequence, one finds the usual temperature $\beta=\frac{2\pi}{\kappa}$. Indeed, the following result 
is easily obtained from the above calculations: 
$
|N_s|^2 =  \frac{1}{e^{\beta \omega}-1},
$
which also gives the mean number of created pairs.\\
In the general case, one proceeds as above, with the only difference that (\ref{sol-gen}) 
is replaced by the more involved expression of the complete solution, where both 
the hypergeometric function and also the more involved factors $(t\pm1)^{l_\pm}$  appear. 
Still, (\ref{straddle}) remains true, and we get 
\beq
|N_s|^2 =  \frac{1}{e^{\beta (\omega +e \Phi_H)}-1},
\eeq
which is expected. 

In order to corroborate the above scattering picture, we can proceed as follows. Let us consider 
the balance of fluxes by using the conserved current 
$
J_r :=-\frac{i}{2} \left(\eta^\ast (\partial_r \eta)-(\partial_r \eta^\ast) \eta\right),
$
in Kruskal-like coordinates, where for the region outside the black hole horizon we get 
$
\partial_r =  U\partial_U +V\partial_V.
$
It is also interesting to point out that, in the inner part of the black hole, where $U>0,V>0$, 
by maintaining the same definition for $u,v$, we have $\partial_r =  -U\partial_U +V\partial_V$, and also 
$\partial_\psi=U \partial_U + V \partial_V$.\\
We get, as solution describing the Hawking radiation process for the Klein-Gordon equation both in the black hole region 
and in the external one, 
the following analytic continuation of the solution (\ref{sol-gen}):
\beq
\eta(U,V)_{hawking}=\left\{
\begin{array}{lr}
c_1 (U)^{i \omega} + c_2 V^{i \omega} &  \textrm{for}\ U>0,\\
c_1 e^{-\pi \omega} (-U)^{i \omega} &  \textrm{for}\ U<0,
\end{array}
\right. 
\eeq
which represents a state composed by an ingoing negative norm state $V^{i \omega}$ and a negative norm 
outgoing state in the black hole region $U>0$ and a outgoing positive norm particle state $e^{-\pi \omega} (-U)^{i \omega}$. 
As to the normalization, we get (cf. also \cite{massar-parentani})
\eqn
c_1&=& \frac{1}{\sqrt{4 \pi \omega}} \frac{1}{\sqrt{1-e^{-2\pi \omega}}},\\
c_2&=& \frac{1}{\sqrt{4 \pi \omega}} .
\feqn
In particular, for the external region, we get external part of the straddling mode. 
Then we find the following transmission coefficient:
\beq
|T|^2=\left| \frac{J_r^{transmitted}}{J_r^{incident}} \right|=\frac{1}{e^{2\pi \omega}-1},
\eeq
which gives again a thermal spectrum 
for created pairs.\footnote{ One may wonder if the flux of $J_r$, which is meaningful in the 
external region $U<0$, is still meaningful also in the black hole region $U>0$. 
We observe that the current involves substantially Wronskian relations also in the 
inner region, and so is conserved also there, even if its physical interpretation is   
not perspicuous.}\\ 
The picture is not strictly the same as in the tunneling ansatz in the HJ formalism, because 
an ingoing internal mode appears too. In particular, we have an antiparticle (negative norm) state 
which, in a scattering picture, is composed by an ingoing negative energy  state traveling backward in time 
towards the interior region (and then an antiparticle traveling forward in time towards the horizon) 
as initial state and a pair composed by a negative norm state  traveling backward in time towards the 
horizon (and then an antiparticle traveling forward in time towards the interior region), and a particle 
state in the external region moving away from the horizon. The last two states are the same as in 
the original picture by Damour and Ruffini \cite{damour-ruffini}. 
We stress that the bonus of the Nariai geometry 
consists in the fact that solutions inside the black hole do not suffer the problem to 
deal with a curvature singularity, and are exact.

\section{Conclusions}

We have discussed some aspects of the tunneling approach to Hawking radiation. In particular, we have shown that  
$\Gamma_{emission}$ can be reinterpreted as a conditional probability of pair emission from vacuum, in the so-called 
transmission coefficient approach \cite{nikishov,damour-mg1}, and that 
the current and expected decay rate for the vacuum is obtained for the emission of thermal 
particles by the black hole. Then, we have studied Hawking emission in the Nariai case, 
with the aim of corroborating the tunneling 
ansatz by exploiting exact solutions of the field equations, which allow to set up an S-matrix approach 
even in the static situation for the Hawking process.  A suitable use of fluxes allows to get the same result than in the semiclassical 
tunneling approach, 
which has been explored in regular (Kruskal-like) coordinate patches, both at the 
black hole horizon and also at the cosmological horizon and extended to the emission of charged particles. \\

\end{document}